\documentclass{elsart} 
\usepackage{graphics} 
\usepackage{bm} 

\begin{document}

\begin{frontmatter} 

\title{Hypernuclear physics legacy and heritage of 
Dick Dalitz\thanksref{ref}} 
\thanks[ref]{Updated, extended version of a plenary talk at HYP06, Mainz, 
October 2006 [A. Gal, in: J.~Pochodzalla, Th.~Walcher (Eds.), Hypernuclear 
and Strange Particle Physics, HYP2006, SIF and Springer, Berlin Heidelberg 
2007, pp. 11-16] published in a special issue on Recent Advances in 
Strangeness Nuclear Physics, A.~Gal and R.S.~Hayano (Eds.), Nucl. Phys. A 
{\bf 804} (2008) 13-24.}

\author{Avraham Gal}        
\ead{avragal@vms.huji.ac.il}
\address{Racah Institute of Physics, The Hebrew University,
Jerusalem 91904, Israel}


\begin{abstract} 
The major contributions of Richard H. Dalitz to hypernuclear 
physics, since his first paper in 1955 to his last one in 2005 covering 
a span of 50 years during which he founded and led the theoretical study 
of hypernuclei, are reviewed from a personal perspective. 
Topical remarks on the search for quasi-bound $\bar K$-nuclear states 
and on kaon condensation are made.  
\end{abstract} 


\end{frontmatter}

\section{Introduction} 
\label{sec:intro} 

Dick Dalitz was born in Dimboola, in the state of Victoria, Australia, 
on February 28th 1925, and gained B.A. and B.Sc. degrees
in Mathematics and Physics in 1944 and 1945, respectively, from the
University of Melbourne. He moved to Britain in 1946 for postgraduate studies 
at Cambridge, and then worked at the University of Bristol before joining in 
1949 Rudolf Peierls in Birmingham. There he completed and wrote up his Ph.D. 
thesis on `$0^+ \to 0^+$ transitions in nuclei', supervised by Nicholas Kemmer 
of Cambridge, and subsequently became a Lecturer. He spent
two years in the U.S. from 1953, holding research positions at Cornell and
Stanford, visiting also Princeton and Brookhaven National Laboratory, 
and returned as a Reader in Mathematical Physics to the University of
Birmingham for a year before becoming Professor of Physics in the Enrico Fermi
Institute for Nuclear Studies and the Department of Physics at the University
of Chicago in 1956. He moved to Oxford in 1963 as a Royal Society
Research Professor, the post he held until his retirement in 1990. 
In addition to the Dalitz Plot, Dalitz Pair and the Castillejo-Dalitz-Dyson 
(CDD) Pole that bear his name, he pioneered the theoretical study of 
strange baryon resonances, of baryon spectroscopy in the quark model, and 
of hypernuclei, to all of which he made outstanding contributions. 
His formulation of the $\theta-\tau$ puzzle led to the discovery 
that parity is not a symmetry of the weak interactions. 
A complete bibliography of Dalitz's works is available in Ref.~\cite{ACG06}. 

During his postgraduate studies he spent a year working alongside Cecil 
Powell's cosmic ray group at Bristol and it was during this period that he 
took particular interest in the strange particles that were beginning to 
appear in cosmic rays and at particle accelerators. These included the first 
hyperfragment in 1952~\cite{DPn53} which inspired a lifelong interest in 
hypernuclei. Later on, he made significant contributions to the strong 
interactions of the strange particles and their resonant 
states~\cite{Dal61,Dal63}. 
As early as 1959 Dalitz and Tuan, by analysing the data on the strong 
interactions of $K^-$ mesons with protons, predicted the existence of an 
$I=0,~J^{\pi}=(1/2)^-$ strange resonance about 20~MeV below the $K^-p$ 
threshold~\cite{Dal59}. This $\Lambda(1405)$ resonance was discovered two 
years later in the Berkeley hydrogen bubble chamber, studying the reaction 
$K^-p \to \Sigma + 3\pi$ for several charge states~\cite{Als61}. 
The proximity of this $s$-wave $\pi \Sigma$ resonance to the $\bar K N$ 
threshold suggested that it can be generated by $\bar K N-\pi\Sigma$ 
inter-hadron forces, and this was shown in 1967 by Dalitz to be possible 
within a dynamical model of SU(3)-octet vector-meson exchange~\cite{DWR67} 
which is, in fact, the underlying physical mechanism for the 
Tomozawa-Weinberg leading term in the chiral expansion of the meson-baryon 
Lagrangian~\cite{Tom66,Wei66}. The vector mesons $\rho, \omega, K^{\star}, 
\phi$, which were discovered in the years 1960-62, relying heavily on Dalitz 
plots for some of these, were unknown when the $\Lambda(1405)$ was predicted. 
In the years to follow, Dalitz repeatedly considered the completeness
of this dynamical picture, whether or not the $S$-matrix pole of 
$\Lambda(1405)$ due to the inter-hadron forces need not be augmented by 
a CDD pole arising from inter-quark forces upon allowing for an intermediate 
$uds$ configuration. It is here that the earlier CDD discussion~\cite{CDD56}
found a fertile physical ground. 

Looking back years later at the development of his own career, he made the 
following remarks~\cite{Dal82} 
(which he rarely allowed himself to make in public): 
\begin{itemize} 
\item 
{\it Yes, as Gell-Mann said, pion physics was indeed the central topic for 
theoretical physics in the mid 1950s, and that was what the young theoretician 
was expected to work on. The strange particles were considered generally to be 
an obscure and uncertain area of phenomena, as some kind of dirt effect which 
could not have much role to play in the nuclear forces, whose comprehension 
was considered to be the purpose of our research. Gell-Mann remarked that he 
spent the major part of his effort on pion physics in that period, and I did 
the same, although with much less success, of course.} 
\item 
{\it Fashions have always been strong in theoretical physics, and that holds 
true today as much as ever. The young physicist who is not working on those 
problems considered central and promising at the time, is at a disadvantage 
when he seeks a post. This tendency stems from human nature, of course, 
but it is unfortunate, I think, that the system operates in such a way as to 
discourage the young physicist from following an independent line of thought.} 
\end{itemize} 
Although about $30\%$ of his research papers were devoted or connected to 
hypernuclei, Dalitz was primarily a particle physicist. This is reflected 
in the interview he gave during HYP03~\cite{Dal04}, where hypernuclei get 
only the following two brief remarks: 
\begin{itemize}
\item
{\it My interest in hypernuclear events developed particularly well in Chicago 
because a young emulsion experimenter, Riccardo Levi-Setti, whose work I had 
known from his hypernuclear studies in Milan, came to the Institute for 
Nuclear Studies at this time. We each benefited from the other, I think, and 
we got quite a lot done.} 
\item 
{\it I was responsible for organizing particle-physics theory in Oxford. 
Besides quark-model work, I still did work on hypernuclear physics, 
much of this with Avraham Gal of Jerusalem.} 
\end{itemize}   

I first met Dalitz as a young student attending the 1966 Varenna 
International School of Physics ``Enrico Fermi", Course XXXVIII on 
`Interaction of High-Energy Particles with Nuclei'. He gave a series of 
lectures on the status of Hypernuclear Physics, and I was lucky to have 
been able to intercept him during one of the lectures, apprising him of 
an important omission he had made in a calculation of transition matrix 
elements with which I was familiar owing to my shell-model education at 
the Weizmann Institute. This was the beginning of a very close collaboration 
lasting about 20 years during which we would often meet for joint periods 
of work, always discussing the latest experimental results and their likely 
interpretations. I have been amazed at Dalitz's encyclopaedic knowledge and 
mastery of measurements and calculations in particle physics and also of 
many aspects of nuclear physics, his critical assessment of experimental 
results and his thoroughness at work. He always insisted on and managed 
to calculate things in his own way, relying only on facts, never on fancy. 
Our ways somewhat diverged after 1985, but we still maintained a close 
relationship until very recently, when I edited his last publication, 
the talk he gave at HYP03~\cite{Dal05}.

\section{$\Lambda$ hypernuclei}
\label{lambda}

\subsection{The beginning} 

Dalitz pioneered the theoretical study of hypernuclei. His first published 
work on $\Lambda$ hypernuclei dates back to 1955, titled {\it Charge 
independence in light hyperfragments}~\cite{Dal55}. It focused on the near 
equality of the ($_{\Lambda}^4{\rm H},~_{\Lambda}^4{\rm He}$) binding energies 
and its origin in the charge symmetry of the $\Lambda N$ interaction, and on 
the exceedingly small binding energy of $_{\Lambda}^3{\rm H}$, the only bound 
$A=3$ hypernucleus marking the onset of $\Lambda$-hypernuclear binding. 
By 1959 his analyses of the light, $s$-shell hyperfragments led him to 
state~\cite{DDa59} that {\it the existence of a bound $\Lambda$-nucleon 
system is strongly excluded} and that {\it the analysis of the $T=1$ triplet 
$_{\Lambda}^3{\rm He},~_{\Lambda}^3{\rm H},~_{\Lambda}^3{\rm n}$ indicates 
that these systems are not expected to form bound states}, and that these 
essential conclusions {\it would not be seriously affected if there exist 
moderately strong three-body forces arising from pion exchange processes.} 
He returned in 1972 to consider the possible effects of three-body 
$\Lambda NN$ forces in the $s$ shell~\cite{DHT72} quantifying what has been 
since called `the overbinding problem', namely that the binding energy of 
$_{\Lambda}^5{\rm He}$ comes out too large by $2-3$~MeV in any calculation 
that fits well the binding energies of the lighter hypernuclei.\footnote{we 
now know that it need not be the case once $\Lambda N - \Sigma N$ coupling 
is explicitly allowed in.} 

In a series of works covering three decades, he used the main 
$\Lambda \to p \pi^-$ weak-decay mode of light hypernuclear species studied 
in emulsion and bubble chambers to determine their ground-state spins and, 
thereby, to gain information on the spin dependence of the $\Lambda N$ force. 
When he had begun this line of works, 
just before parity violation was realised during the turbulent 1956-1957 
period, he wrongly concluded in a talk given at the 6th Annual Rochester 
Conference on High Energy Nuclear Physics in April 1956 that the triplet 
$\Lambda N$ $s$-wave interaction was stronger than the singlet 
one~\cite{Dal56}. His argument was based on assuming that parity was 
respected in the weak decay $_{\Lambda}^4{\rm H} \to \pi^- + {^4{\rm He}}$. 
Since the final products all had spin zero, and the pion was known to have 
a negative intrinsic parity with respect to nucleons, (quoting Dalitz, in 
{\it italics}) {\it the spin-parity 
possibilities for the ($_{\Lambda}^4{\rm H},~_{\Lambda}^4{\rm He}$) doublet 
are $0^-,~1^+,~2^-$, etc.} Assuming (at that time it was still uncertain) 
that the $\Lambda$ hyperon had spin-parity $(1/2)^+$, the spin-parity of 
$_{\Lambda}^4{\rm H}$ had to be $1^+$, and this meant that the triplet 
$\Lambda N$ $s$-wave interaction was stronger than the singlet one, and 
{\it one also concludes that the spin-parity for $_{\Lambda}^3{\rm H}$ is 
$(3/2)^+$.} Of course we now know that this was wrong; and indeed soon after 
Dalitz himself, realising the merits of the strong spin selectivity provided 
by parity violation in the weak-interaction pionic decays of $\Lambda$ 
hypernuclei, calculated the branching ratios of the 
$\pi^-$ two-body decays of $_{\Lambda}^4{\rm H}$ and $_{\Lambda}^3{\rm H}$ 
to the daughter ground states of $^4{\rm He}$ and $^3{\rm He}$, respectively, 
in order to determine unambiguously the ground-state spins of the parent 
hypernuclei~\cite{DLi59} which in a few years became experimentally 
established as $0^+$~\cite{ALS61} and $(1/2)^+$~\cite{ADH62} respectively. 
This led to the correct ordering of the triplet and singlet $\Lambda N$ 
$s$-wave interactions as we understand it to date. 

Dalitz's outstanding contribution in the 1960s to weak interactions in 
hypernuclei, together with Martin Block \cite{BDa63}, was to formulate the 
$\Lambda N \to NN$ phenomenology of non-mesonic weak-interaction decay modes 
that dominate the decays of medium-weight and heavy hypernuclei, 
a process that cannot be studied on free baryons and which offers new systems, 
$\Lambda$ hypernuclei, for exploring the little understood $\Delta I = 1/2$
rule in non-leptonic weak interactions. The state of the art in understanding 
non-mesonic weak decay in $\Lambda$ hypernuclei is reviewed in this Issue, 
experimentally by H.~Outa and theoretically by G.~Garbarino. This chapter 
in hypernuclear physics is still incomplete, and more experimentation is 
needed before the underlying physics is fully uncovered.

\subsection{The later years} 

Dalitz's work on the $p$-shell hypernuclei, dates back to 1963 when 
together with Levi Setti, in their only joint paper~\cite{DLe63}, 
{\it Some possibilities for unusual light hypernuclei} were discussed, 
notably the neutron-rich isotopes of $_{\Lambda}^6{\rm H}$ and 
$_{\Lambda}^8{\rm He}$ belonging to $I=3/2$ multiplets, but his systematic 
research of the $p$-shell hypernuclei started in 1967 together with me 
laying the foundations for a shell-model analysis of $\Lambda$ hypernuclei 
\cite{GSD71}. Using these shell-model techniques, subsequently we were able 
to chart the production and $\gamma$-ray decay schemes anticipated for 
excited states in light $\Lambda$ hypernuclei in order to derive the 
complete spin dependence of the $\Lambda N$ interaction effective in 
these hypernuclei~\cite{DGa78}. This work was further developed together 
with John Millener and Carl Dover~\cite{MGD85}, serving as a useful guide 
to the hypernuclear $\gamma$-ray measurements completed in the 
last few years, at BNL and at KEK~\cite{HTa06}, which yielded full 
determination of the spin dependence in the low-lying spectrum 
(contributions by H.~Tamura and by D.J.~Millener, in this Issue). 

As early as 1969 data on excited states were reported with the $\Lambda$ 
hyperon in a $(1p)_{\Lambda}$ state coupled to the nuclear ground-state 
configuration, first from emulsion data~\cite{DSa69,Boh70} observing proton 
decay in some special instances such as $_{\Lambda}^{12}{\rm C}$, and later 
on through in-flight $(K^-,\pi^-)$ experiments at CERN and BNL. In the 
particular case of the $_{\Lambda}^{12}{\rm C}$ excited cluster of states 
about 11 MeV above the $(1s)_{\Lambda}$ ground state, Dalitz participated 
actively in the first round of theoretical analysis for both types of 
experiments~\cite{DDT86,DGW79}. However, confronting these and similar data 
posed two difficulties which we identified and discussed during 1976. The 
first one was connected to understanding the nature of the $\Lambda$ continuum 
spectrum which, owing to the small momentum transfer in the forward-direction 
$(K^-,\pi^-)$ reaction in flight, was thought to consist of well defined 
$\Lambda$-hypernuclear excitations. It was not immediately recognised that 
since the $\Lambda$ hyperon did not have to obey the Pauli exclusion principle 
with nucleons, hypernuclear quasi-free excitation was possible even at 
extremely small values of the momentum transfer, a possibility that was 
pointed out and analysed quantitatively by us~\cite{DGa76a} following the 
first round of data taken by the Heidelberg-Saclay collaboration at the 
CERN-PS in 1975. 

The other difficulty was connected with understanding the role of coherent 
excitations in the $(1p)_{\Lambda}$ continuum, the so called `substitutional' 
or `analogue' states, where the early theoretical concept of analogue states 
stemmed from considerations of octet-SU(3) unitary symmetry. 
Already in his first discussion of these states in 1969~\cite{Dal69}, 
Dalitz recognised {\it that the strong excitation of these states 
does not depend on SU(3) symmetry. In fact it is reasonable to believe that 
SU(3) symmetry has almost no relevance to the relationship between 
$\Lambda$-hypernuclei and nuclei...simply because the mass difference of 
80 MeV between the $\Lambda$ and $\Sigma$ hyperons...is a very large 
energy relative to the typical energies associated with nuclear excitations.} 
This difficulty was eliminated by Kerman and Lipkin~\cite{KLi71} who suggested 
in 1971 to consider the Sakata triplet-SU(3) unitary symmetry version in which 
the proton, neutron and $\Lambda$ were degenerate. This suggestion was further 
limited by us in 1976 to $(1p)_{p,n,\Lambda}$ states and, together with 
Pauli-spin SU(2) symmetry, led to the consideration of Pauli-Sakata SU(6) 
supermultiplets encompassing nuclei and hypernuclei~\cite{DGa76b}, in direct 
generalisation of Wigner's supermultiplet theory of spin-isospin SU(4) 
symmetry in light nuclei. The analysis of these SU(6) supermultiplets proved 
very useful for the development of shell model techniques in the 1980s 
and on by John Millener and collaborators~\cite{ABD81}. 
In particular, the 1976 work focused on the concept of the `supersymmetric' 
state in addition to the `analogue' state, with the low-lying 
supersymmetric state arising from the non existence of a Pauli exclusion 
principle between the $\Lambda$ hyperon and nucleons.

\section{$\Lambda\Lambda$ hypernuclei}
\label{sec:lambdalambda} 

Dalitz in fact anticipated that $\Lambda\Lambda$ hypernuclei be observed and 
that as a rule they would be particle stable with respect to the strong 
interaction. His Letter titled {\it The $\Lambda\Lambda$-hypernucleus 
and the $\Lambda-\Lambda$ interaction}~\cite{Dal63a} appeared as soon as the 
news of the first observed $\Lambda\Lambda$-hypernucleus 
$_{\Lambda\Lambda}^{10}{\rm Be}$ was reported in 1963~\cite{DGP63} 
and was followed by a regular paper~\cite{DRa64}. He did not work on 
$\Lambda\Lambda$ hypernuclei for a long period, until 1989, apparently 
because there were no new experimental developments in this field except 
for the $_{\Lambda\Lambda}^{~6}{\rm He}$ dubious event reported by Prowse in 
1966. He returned to this subject in 1989~\cite{DDF89} feeling the need to 
scrutinize carefully the interpretation of the 
$_{\Lambda\Lambda}^{10}{\rm Be}$ event and its implications in view of 
a renewed experimental interest to search for the $H$-dibaryon. This 
scientific chapter in Dalitz's life is described in Don Davis' companion 
contribution in this Issue.

\section{$\Sigma$ hypernuclei} 
\label{sec:sigma} 

Dalitz was puzzled by the CERN-PS low-statistics evidence in the beginning 
of the 1980, and subsequently by the KEK-PS low-statistics evidence in 1985, 
for relatively narrow $\Sigma$-hypernuclear peaks in the continuum. The large 
$\Sigma N \to \Lambda N$ low-energy cross section, due primarily to the 
strong pion exchange potential, did not leave much room for narrow $\Sigma$ 
states in nuclei; indeed, the first rough estimate by Gal and 
Dover~\cite{GDo80} gave nuclear-matter widths of order 
$\Gamma_{\Sigma} \sim 25$~MeV. The suggestion by these authors that some 
$\Sigma$-hypernuclear levels could selectively become fairly narrow due 
to the $S=1,~I=1/2$ dominance of the $\Sigma N \to \Lambda N$ transition 
fascinated him to the extent that he argued favorably for the validity of 
this interpretation in his 1980 Nature article {\it Discrete 
$\Sigma$-hypernuclear states}~\cite{Dal80}, although taking it with a grain 
of salt. He came back to this subject in 1989, after hearing in HYP88 at 
Padova Hayano's report of the KEK $K^-_{\rm stop}$ experiment \cite{HII89} 
finding evidence for a $_{\Sigma}^4{\rm He}$ near-threshold narrow state. 
Recalling some old bubble-chamber data on $K^-_{\rm stop}$ absorption 
yields in $^4$He near the $\Sigma$ threshold, he questioned together 
with Davis and Deloff~\cite{DDD90} the compatibility of assigning this 
$_{\Sigma}^4{\rm He}$ as a quasi-bound state with the older data: 
{\it Is there a bound $_{\Sigma}^4{\rm He}$?} 
He came back to these questions with Deloff in both HYP91 in Shimoda 
\cite{DDe92} and HYP94 in Vancouver \cite{DDe95}, but as soon as this same 
$_{\Sigma}^4{\rm He}$ structure was observed in a ($K^-,\pi^-$) in-flight 
experiment at the BNL-AGS accelerator \cite{NMF98}, and the measured pion 
spectrum was explained satisfactorily within a comprehensive DWIA calculation 
\cite{Har98} in terms of a $_{\Sigma}^4{\rm He}$ quasi-bound state, he openly 
during HYP00 \cite{Dal01} and HYP03 \cite{Dal05} removed his objections.

\section{Exotic structures} 
\label{sec:exotic} 

I have already mentioned that Dalitz was far from jumping on band wagons of 
speculative ideas unless there were some good experimental or phenomenological 
tests to be made in a concrete manner. A notable exception is a Nature paper 
coauthored by Dalitz, {\it Growing drops of strange matter}~\cite{SSD89}, 
discussing a possible scenario for getting into strange quark matter. It is 
therefore timely to wonder how Dalitz would have reacted to the flood of 
recent reports on the possible existence of $\bar K$-nuclear bound states 
and on the ongoing experimental searches for such objects, particularly 
since the prototype of such states, the $\Lambda(1405)$, was first interpreted 
and calculated in a paper led by him \cite{DWR67} as an unstable $\bar K N$ 
bound state.

\subsection{$\bar K$-nuclear bound states?} 

The state of the art in searches for $\bar K$-nuclear bound states is 
reviewed in this Issue by M.~Iwasaki for the KEK experiments, stopping $K^-$ 
mesons on a $^4$He target, and by A.~Filippi for the FINUDA spectrometer 
collaboration in Frascati, stopping $K^-$ mesons on several targets, including 
isotopes of Li and $^{12}$C. At the background of these stopped $K^-$ 
experiments was the prediction of a particularly narrow and deeply bound $I=0$ 
${\bar K} NNN$ tribaryon, and the related estimate given by Akaishi and 
Yamazaki for its production rate $\sim 2 \%$ per stopped $K^-$ in $^4{\rm He}$ 
\cite{AYa02}. This estimate is totally unacceptable since a production rate of 
this order of magnitude is known to hold at rest for (the most favourable) 
$A=4$ hypernuclei~\cite{Dav05}; hypernuclei are produced via the dominant 
absorptive $K^- N \to \pi Y$ modes, whereas the $K^- N \to N \bar K$ 
backward-elastic mode responsible for replacing a bound nucleon by a bound 
$\bar K$ is suppressed at rest with respect to the former reactive modes 
owing to the $1/v$ law near threshold. Realistic estimates give rates as low 
as $10^{-4}$ per stopped $K^-$ for the production of $\bar K$-nuclear bound 
states \cite{Gal08}. Indeed, an upper limit of order $10^{-3}$ per stopped 
$K^-$ for producing narrow ${\bar K} NNN$ deeply bound states has been 
determined recently by the KEK-E549 collaboration from the observed 
$(K^{-}_{\rm stop},p)$ spectrum in $K^-$ absorption on $^4$He \cite{SBC07a}. 
In-flight $K^-$ reactions are more promising, but unfortunately will not be 
feasible before J-PARC is operated, from 2009 on. Preliminary ($K^-,p$) and 
($K^-,n$) spectra at $p_{\rm lab}=1$~GeV/c on $^{12}$C obtained in KEK-E548 
show only appreciable strength in the $\bar K$ bound-state region, but no 
peaks~\cite{KHA07}, in accordance with recent in-flight reaction calculations 
(Ref.~\cite{YNH06}; T.~Koike and T.~Harada, in this Issue). 
Given this situation, the measurement of more exclusive spectra, and using 
proton or antiproton beams, or nucleus-nucleus collisions, has been advocated 
(P.~Kienle, in this Issue). 

\begin{figure}[t] 
\begin{center} 
\resizebox{1.0\textwidth}{!}{%
\includegraphics{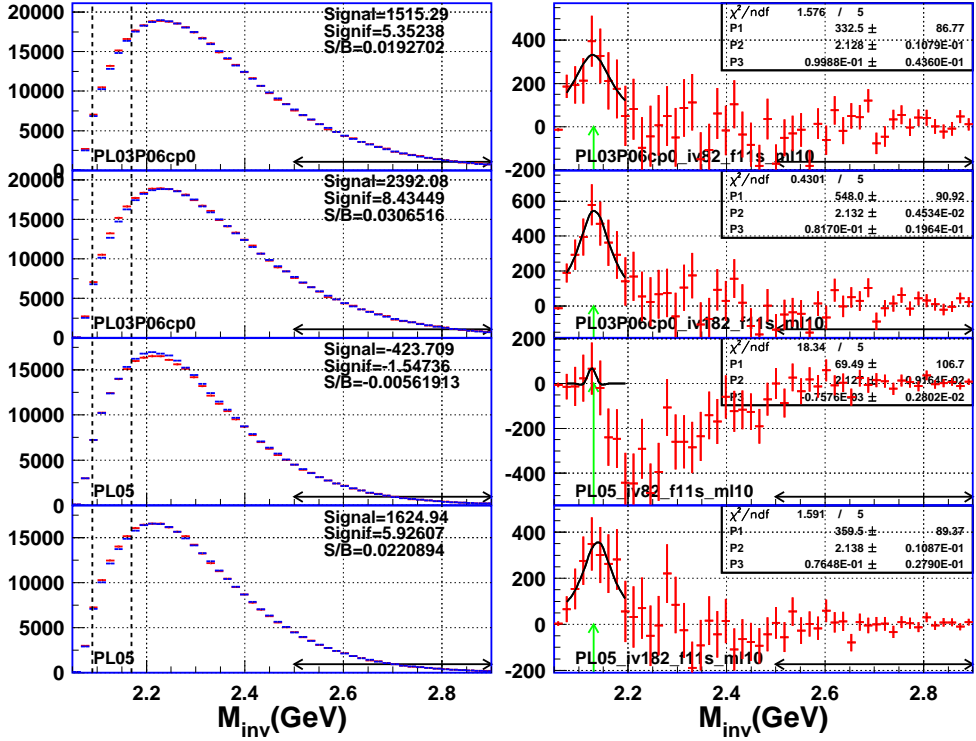}} 
\end{center} 
\caption{$\Lambda p$ invariant-mass spectra taken by the FOPI detector 
collaboration at GSI in Ni+Ni (two upper panels) and in Al+Al (two lower 
panels) collisions. The right-hand side panels follow alignment of the 
reaction plane (upper panel in each group) or alignment of the $\Lambda$ 
direction (lower panel in each group). Figure provided by Norbert Herrmann 
and shown by Paul Kienle at HYP06 and in his contribution to this Issue 
\cite{Kie06}. I am indebted to both of them for bringing these data to my 
attention and for instructive discussions.} 
\label{fig:kienle} 
\end{figure} 

Among the many reactions and spectra presented and discussed recently, 
I would like to propose interpretation to a class of very intriguing spectra 
that naively would be interpreted as due to an extremely deep $K^-pp$ bound 
state. Preliminary results from the FOPI collaboration at GSI are shown in 
Fig.~\ref{fig:kienle}, where the $\Lambda p$ invariant mass in both Ni+Ni 
and Al+Al collisions peaks at $M_{\rm inv}(\Lambda p)=2.13 \pm 0.02$~GeV, 
near the $\Sigma N$ threshold, with an appreciable width. This value of 
$M_{\rm inv}(\Lambda p)$ is substantially lower, by over 100 MeV, than the 
$M_{\rm inv}(\Lambda p)$ value assigned by the FINUDA spectrometer 
collaboration~\cite{ABB05} as due to a $K^-pp$ bound state. The possibility 
of a resonance or cusp phenomenon for the $\Lambda p$ system, at or near the 
opening of the $\Sigma N$ threshold, which has been suggested in several old 
experiments~\cite{Tan69,PBC85}, has always intrigued Dalitz who together 
with others considered it within $K^-d$ calculations~\cite{DHM80,TDD86}, 
in parallel to the Faddeev calculations done by my Ph.D. student Gregory 
Toker~\cite{TGE79}. However, I dare say that had he been with us today, 
he would have considered favourably another possibility, that the light, 
only $\Sigma$ hypernucleus known to be bound, $_{\Sigma}^4{\rm He}$ is the 
source of these $\Lambda p$ pairs. The binding energy of $_{\Sigma}^4{\rm He}$ 
with respect to the $\Sigma^+ + {^3{\rm H}}$ threshold is 
$B = 4.4 \pm 0.3({\rm stat}) \pm 1({\rm syst})$~MeV, and the value of 
width assigned to it is $\Gamma = 7.0 \pm 0.7 + 1.2$~MeV~\cite{NMF98}. 
Its quantum numbers are $I=1/2, J^{\pi}=0^+$~\cite{Har98} with all four 
baryons in $s$ states. In particular, it may be viewed in isospace as a linear 
combination of $\Sigma^+$ coupled to $^3$H and $\Sigma^0$ coupled to $^3$He. 
Its wavefunction is schematically given by: 
\begin{equation} 
\Psi(_{\Sigma}^4{\rm He}) = \alpha {(\Sigma N)^{S=0}_{I=1/2,3/2}} 
{(NN)^{S=0}_{I=1}} + \beta {(\Sigma N)^{S=1}_{I=1/2}} {(NN)^{S=1}_{I=0}}~,
\label{wf} 
\end{equation} 
where only the spin-isospin structure is specified. In the absence of 
dynamical correlations, spin and isospin considerations yield values 
$\alpha^2=\beta^2=1/2$. The decay of $_{\Sigma}^4{\rm He}$ is dominated 
by the ${(\Sigma N \to \Lambda N)^{S=1}_{I=1/2}}$ two-body transition, 
proceeding therefore through the component with amplitude $\beta$ in which 
the $NN$ composition is $pn$. This means that the $\Sigma N$ composition is 
a mixture of $\Sigma^+ n$ and $\Sigma^0 p$, both of which decay to 
$\Lambda p$. One expects then $_{\Sigma}^4{\rm He}$ to decay dominantly by 
emitting back-to-back $\Lambda p$ pairs with slower `spectator' proton and 
neutron which will somewhat distort the $\Sigma N \to \Lambda p$ two-body 
kinematics. A more conclusive proof for this suggestion would come from the 
observation of back-to-back $\Lambda ^3{\rm He}$ pairs in the two-body decay 
$_{\Sigma}^4{\rm He} \to \Lambda + {^3{\rm He}}$. The branching ratio for 
this decay relative to the inclusive $\Lambda X$ decay rate is perhaps 
a few percent, as may be argued by analogy with the approximately 
$8\%(5\%)$ branching ratio measured for the nonmesonic decay  
$_{\Lambda}^4{\rm He} (_{\Lambda}^5{\rm He}) \to n + {^3{\rm He}} 
(^4{\rm He})$ relative to the inclusive $\pi^-$ decay rate of 
$_{\Lambda}^4{\rm He} (_{\Lambda}^5{\rm He})$~\cite{CSO70,KSW76}. 
Irrespective of whether or not the above conjecture of $_{\Sigma}^4{\rm He}$ 
production is correct for the FOPI-Detector GSI experiments, it would be 
a wise practice for $\bar K$-nuclear bound state searches in heavy ion 
collisions to look first for known hypernuclear signals in order to 
determine their production rates as calibration and normalization standards.  

\begin{table} 
\begin{center} 
\caption{Binding energies ($B$) and widths ($\Gamma$) calculated for 
$K^-pp$ (in MeV) exclusive of $\bar K NN \to YN$ contributions} 
\label{tab:kpp} 
\begin{tabular}{lccccc} 
\hline 
 &\multicolumn{2}{c}{single channel} &\multicolumn{2}{c}
{coupled channels} & experiment \\ 
 & ATMS~\cite{YAk02} & AMD~\cite{DHW08} & Faddeev~\cite{SGM07}& 
Faddeev~\cite{ISa07} & FINUDA~\cite{ABB05}\\ \hline 
$B$  & 48 & 16--22 & 50--70 & 60--95 & 115 $\pm$ 6 $\pm$ 4\\ 
$\Gamma$ & 61 & 40--70  & 90--110 & 45--80 & 67 $\pm$ 14 $\pm$ 3\\ 
\hline 
\end{tabular}  
\end{center} 
\end{table} 

\subsection{From $K^-pp$ to kaon condensation?} 

\begin{figure} 
\begin{center} 
\resizebox{0.7\textwidth}{!}{%
\includegraphics{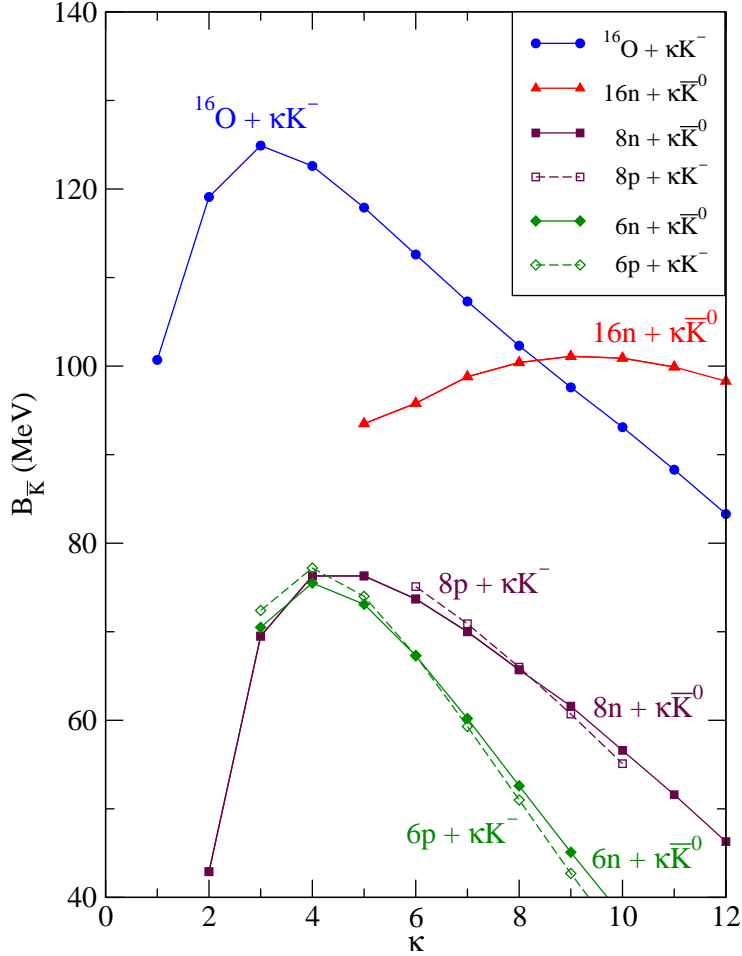}} 
\end{center} 
\caption{1s $\bar K$ separation energy $B_{\bar K}$ in multi-${\bar K}$ 
nuclear systems, as a function of the number of ${\bar K}$ mesons $\kappa$, 
calculated in the NL-SH RMF model \cite{SNR93}, with ${\bar K}$ vector 
coupling constants corresponding to the Tomozawa-Weinberg lowest order chiral 
Lagrangian \cite{Tom66,Wei66} augmented by a ${\bar K}$ scalar coupling 
designed to yield $B_{\bar K}=100$ MeV for $^{16}{\rm O}+1K^-$. 
I thank Daniel Gazda and Ji\v{r}\'{i} Mare\v{s} for providing this figure.} 
\label{fig:multex} 
\end{figure} 

Of special interest is the lighteset $\bar K$-nucleus $K^-pp$. 
The FINUDA spectrometer collaboration claimed evidence for a broad 
$K^-pp$ bound state, with binding energy of about 115 MeV, by observing 
back-to-back $\Lambda p$ pairs from the decay $K^-pp \to \Lambda p$ in 
$K^-_{\rm stop}$ reactions on Li and $^{12}$C (Ref.~\cite{ABB05}; 
A.~Filippi, in this Issue). However, these pairs could naturally arise from 
absorption reactions at rest when final-state interaction is accounted for 
(A.~Ramos, in this Issue). Irrespective of whether or not the FINUDA event 
corresponds to a $K^-pp$ bound state, a variety of few-body calculations 
as summarized in Table~\ref{tab:kpp} agree that $K^-pp$ should possess 
a broad state bound by less than 100 MeV. Here, $K^-pp$ stands loosely for 
the projection onto the $S=0,~I=1/2,~I_z=1/2$ $s$-wave component of 
$\bar K NN$. Its charge symmetric state, with $I_z=-1/2$, is ${\bar K}^0 nn$. 
The table demonstrates that ${\bar K}$ mesons can bind nuclear clusters that 
are otherwise unbound. The point here is that the underlying $K^-p$ and 
${\bar K}^0 n$ interactions (each with equally mixed $I=0$ and $I=1$ 
components) provide considerably more attraction than the purely $I=1$ $K^-n$ 
and ${\bar K}^0 p$ interactions which also contribute in nuclear matter. 
Recent RMF calculations by Gazda et al. \cite{GFG08} demonstrate that 
a finite number of neutrons (protons) can be made self-bound by adding 
together a few ${\bar K}^0$ ($K^-$) mesons, with ${\bar K}$ separation 
energies reaching values $B_{\bar K} \sim 50-100$~MeV. This is shown in 
Fig.~\ref{fig:multex}, where $B_{\bar K}$ is plotted as a function of the 
number of ${\bar K}$ mesons, $\kappa$, for systems of six, eight and sixteen 
neutrons. Shown also are the charge symmetric systems of six and eight 
protons. These systems are particle stable when ${\bar K}$ charge exchange, 
strangeness exchange and multinucleon absorption channels are switched off, 
as practised in this particular calculation. For comparison, $B_{K^-}$ 
values for multi-$K^-$ nuclei based on $^{16}$O are also shown in the figure. 
The `exotic' stable neutron configurations are more tightly bound than in 
the corresponding ordinary nuclei with $N \approx Z$ along the stability 
valley, and the neutron single-particle spectra display substantial 
rearrangement \cite{GFG08}. However, the total binding energy of these 
exotic systems, with ${\bar K}$ mesons, is lower than that of commensurate 
ordinary nuclei with ${\bar K}$ mesons, so that the exotic systems are 
unstable against decay to ordinary nuclei by multiple charge-exchange 
${\bar K}^0 + n \to K^- + p$ reactions. 

The various curves in Fig.~\ref{fig:multex} show a rise of $B_{\bar K}$ 
with $\kappa$, reaching a maximum value and then decreasing steadily with 
$\kappa$. This decrease, due to the enhanced role of vector-meson repulsion 
among ${\bar K}$ mesons upon increasing $\kappa$, persists in a robust way 
within various mean field models and over a broad range of coupling-constant 
values. The maximum values of $B_{\bar K}$ do not exceed the range of values 
100-200 MeV considered normally as providing deep binding for one antikaon. 
This range of binding energies leaves antikaons in multi-${\bar K}$ nuclei 
comfortably above the range of energies where hyperons are relevant. 
It is unlikely therefore that multi-${\bar K}$ nuclei may offer precursor 
phenomena in nuclei, under laboratory conditions, towards kaon condensation. 
Kaon condensation, however, is not ruled out in neutron stars, where time 
scales of the weak interactions are operative, allowing for example a rare 
weak decay such as $e^- \to K^- + \nu_e$ to transform `dense' electrons into 
antikaons, once the effective mass of $K^-$ mesons drops below 200 MeV 
approximately.

\section{Concluding remarks} 
\label{sec:concl}

Dalitz's lifelong study of hypernuclei was central to his career as 
a phenomenologically inclined theoretical physicist. 
His style was unique. Asked by his then student Chris Llewellyn-Smith 
about `new theories', Dalitz responded  
\begin{itemize} 
\item 
{\it My job is not to make theories - it's to understand the data,} 
\end{itemize} 
{\it he saw the theorist's role as being to find a way of representing 
experimental data so that they directly reveal nature's secrets, as the 
Dalitz Plot had done}~\cite{CLS06}. His lifelong nourishment of hypernuclei 
has shaped and outlined for the last 50 years a field that is now maturing 
into a broader context of Strangeness Nuclear Physics. His wise and critical 
business-like attitude will be missed as new experimental facilities are 
inaugurated with the promise of discovering new facets of this field.


\begin{thebibliography}{99}

\bibitem{ACG06} I.J.R. Aitchison, F.E. Close, A. Gal, D.J. Millener, 
Nucl. Phys. A {\bf 771} (2006) 8. 

\bibitem{DPn53} M. Danysz, J. Pniewski, Phil. Mag. {\bf 44} (1953) 348. 

\bibitem{Dal61} R.H. Dalitz, Rev. Mod. Phys. {\bf 33} (1961) 471. 

\bibitem{Dal63} R.H. Dalitz, Annu. Rev. Nucl. Sci. {\bf 13} (1963) 339. 

\bibitem{Dal59} R.H. Dalitz, S.F. Tuan, Phys. Rev. Lett. {\bf 2} (1959) 425. 

\bibitem{Als61} M. Alston, et al., Phys. Rev. Lett. {\bf 6} (1961) 698. 

\bibitem{DWR67} R.H. Dalitz, T.C. Wong, G. Rajasekaran, Phys. Rev. {\bf 153} 
(1967) 1617. 

\bibitem{Tom66} Y. Tomozawa, Nuovo Cimento A {\bf 46} (1966) 707. 

\bibitem{Wei66} S. Weinberg, Phys. Rev. Lett. {\bf 17} (1966) 616. 

\bibitem{CDD56} L. Castillejo, R.H. Dalitz, F.J. Dyson, Phys. Rev. {\bf 101} 
(1956) 453.

\bibitem{Dal82} R.H. Dalitz, Strange particle theory in the cosmic-ray period, 
in \textit{Proc. Colloque Int. sur l'Histoire de la Physique des Particules, 
Paris 1982}, J. Physique {\bf 43}, Colloq. C{\bf 8} Suppl. (1982) pp.~195-205, 
and comments made on p.~406. 

\bibitem{Dal04} M. O'Byrne, From 'tau' to 'top' - the man behind the 
Dalitz plot, CERN COURIER {\bf 44}, issue 2 (2004) article 19. 

\bibitem{Dal05} R.H. Dalitz, Nucl. Phys. A {\bf 754} (2005) 14c. 

\bibitem{Dal55} R.H. Dalitz, Phys. Rev. {\bf 99} (1955) 1475. 

\bibitem{DDa59} B.W. Downs, R.H. Dalitz, Phys. Rev. {\bf 114} (1959) 593. 

\bibitem{DHT72} R.H. Dalitz, R.C. Herndon, Y.C. Tang, Nucl. Phys. B {\bf 47} 
(1972) 109. 

\bibitem{Dal56} R.H. Dalitz, Nature of the $\Lambda^0-N$ force from binding 
energies of light hyperfragments, in \textit{Proc. 6th Annual Rochester 
Conference on High Energy Nuclear Physics, Rochester 1956}, 
Eds. J.~Ballam, et al. (Interscience, New York, 1956) pp. V 40-43. 

\bibitem{DLi59} R.H. Dalitz, L. Liu, Phys. Rev. {\bf 116} (1959) 1312. 

\bibitem{ALS61} R.~Ammar, R.~Levi~Setti, W.~Slater, S.~Limentani, P.~Schlein, 
P.~Steinberg, Nuovo Cimento {\bf 19} (1961) 20. 

\bibitem{ADH62} R.~Ammar, W.~Dunn, M.~Holland, Nuovo Cimento {\bf 26} (1962) 
840. 

\bibitem{BDa63} M.M. Block, R.H. Dalitz, Phys. Rev. Lett. {\bf 11} (1963) 96. 

\bibitem{DLe63} R.H. Dalitz, R. Levi Setti, Nuovo Cimento {\bf 30} (1963) 489. 

\bibitem{GSD71} A. Gal, J.M. Soper, R.H. Dalitz, Ann. Phys. {\bf 63} (1971) 
53; {\bf 72} (1972) 445; {\bf 113} (1978) 79. 

\bibitem{DGa78} R.H. Dalitz, A. Gal, Ann. Phys. {\bf 116} (1978) 167. 

\bibitem{MGD85} D.J. Millener, A. Gal, C.B. Dover, R.H. Dalitz,
Phys. Rev. C {\bf 31} (1985) 499. 

\bibitem{HTa06} O. Hashimoto, H. Tamura, Prog. Part. Nucl. Phys. {\bf 57} 
(2006) 564. 

\bibitem{DSa69} D.H. Davis, J. Sacton, Hypernuclear physics using photographic 
emulsion, in {\it Proc. Int. Conf. Hypernuclear Physics, ANL 1969}, Eds. 
A.R. Bodmer, L.G. Hyman (ANL, Argonne IL, 1969) pp. 159-192. 

\bibitem{Boh70} G. Bohm, et al., Nucl. Phys. B {\bf 24} (1970) 248; 
M. Juri\'{c}, et al., Nucl. Phys. B {\bf 47} (1972) 36. 

\bibitem{DDT86} R.H. Dalitz, D.H. Davis, D.N. Tovee, Nucl. Phys. A {\bf 450} 
(1986) 311c. 

\bibitem{DGW79} C.B. Dover, A. Gal, G.E. Walker, R.H. Dalitz, Phys. Lett. B 
{\bf 89} (1979) 26. 

\bibitem{DGa76a} R.H. Dalitz, A. Gal, Phys. Lett. B {\bf 64} (1976) 154. 

\bibitem{Dal69} R.H. Dalitz, The present problems and future outlook in 
$\Lambda$-hypernuclear physics, in {\it Proc. Int. Conf. Hypernuclear Physics, 
ANL 1969}, Eds. A.R. Bodmer, L.G. Hyman (ANL, Argonne IL, 1969) pp. 708-747. 

\bibitem{KLi71} A.K. Kerman, H.J. Lipkin, Ann. Phys. {\bf 66} (1971) 738. 

\bibitem{DGa76b} R.H. Dalitz, A. Gal, Phys. Rev. Lett. {\bf 36} (1976) 362.  

\bibitem{ABD81} E.H. Auerbach, A.J. Baltz, C.B. Dover, A. Gal, S.H. Kahana, 
L. Ludeking, D.J. Millener, Phys. Rev. Lett. {\bf 47} (1981) 1110; Ann. Phys. 
{\bf 148} (1983) 381. 

\bibitem{Dal63a} R.H. Dalitz, Phys. Lett. {\bf 5} (1963) 53. 

\bibitem{DGP63} M. Danysz, et al., Phys. Rev. Lett. {\bf 11} (1963) 29; 
Nucl. Phys. {\bf 49} (1963) 121. 

\bibitem{DRa64} R.H. Dalitz, G. Rajasekaran, Nucl. Phys. {\bf 50} (1964) 450. 

\bibitem{DDF89} R.H. Dalitz, D.H. Davis, P.H. Fowler, A. Montwill, 
J.~Pniewski, J.A.~Zakrzewski, Proc. Royal Soc. A {\bf 426} (1989) 1. 

\bibitem{GDo80} A. Gal, C.B. Dover, Phys. Rev. Lett. {\bf 44} (1980) 379. 

\bibitem{Dal80} R.H. Dalitz, Nature {\bf 285} (1980) 11. 

\bibitem{HII89} R.S. Hayano, et al., Nuovo Cimento A {\bf 102} (1989) 
437. 

\bibitem{DDD90} R.H. Dalitz, D.H. Davis, A. Deloff, Phys. Lett. B {\bf 236} 
(1990) 76. 

\bibitem{DDe92} R.H. Dalitz, A. Deloff, Nucl. Phys. A {\bf 547} (1992) 181c.

\bibitem{DDe95} R.H. Dalitz, A. Deloff, Nucl. Phys. A {\bf 585} (1995) 303c.

\bibitem{NMF98} T. Nagae, et al., Phys. Rev. Lett. {\bf 80} (1998) 1605. 

\bibitem{Har98} T. Harada, Phys. Rev. Lett. {\bf 81} (1998) 5287. 

\bibitem{Dal01} R.H. Dalitz, Nucl. Phys. A {\bf 691} (2001) 1c. 

\bibitem{SSD89} G.L. Shaw, M. Shin, R.H. Dalitz, M. Desai, Nature {\bf 337} 
(1989) 436. 

\bibitem{AYa02} Y. Akaishi, T. Yamazaki, Phys. Rev. C {\bf 65} (2002) 044005. 

\bibitem{Dav05} D.H. Davis, Nucl. Phys. A {\bf 754} (2005) 3c. 

\bibitem{Gal08} A.~Gal, in preparation. 

\bibitem{SBC07a} M.~Sato, et al., Phys. Lett. B {\bf 659} (2008) 107. 

\bibitem{KHA07} T. Kishimoto, et al., Prog. Theor. Phys. {\bf 118} 
(2007) 181. 

\bibitem{YNH06} J. Yamagata, H. Nagahiro, S. Hirenzaki, Phys. Rev. C {\bf 74} 
(2006) 014604.  

\bibitem{Kie06} P. Kienle, in {\it Hypernuclear and Strange Particle Physics}, 
Proc. HYP06, Eds. J.~Pochodzalla, Th.~Walcher (SIF and Springer-Verlag, 
Berlin Heidelberg, 2007) pp.~225-229; Nucl. Phys. A {\bf 804} (2008) 286. 

\bibitem{ABB05} M. Agnello, et al., Phys. Rev. Lett. {\bf 94} (2005) 
212303. 

\bibitem{Tan69} T.H. Tan, Phys. Rev. Lett. {\bf 23} (1969) 395. 

\bibitem{PBC85} C. Pigot, et al., Nucl. Phys. B {\bf 249} (1985) 172. 

\bibitem{DHM80} R.H. Dalitz, C.R.~Hemming, E.J.~Morris, Nukleonika {\bf 25} 
(1980) 1555; R.H.~Dalitz, A.~Deloff, Czech J.~Phys. B {\bf 32} (1982) 1021; 
R.H.~Dalitz, A.~Deloff, Australian J.~Phys. {\bf 36} (1983) 617. 

\bibitem{TDD86} M.~Torres, R.H.~Dalitz, A.~Deloff, Phys. Lett. B {\bf 174} 
(1986) 213. 

\bibitem{TGE79} G.~Toker, A.~Gal, J.M.~Eisenberg, Phys. Lett. B {\bf 88} 
(1979) 235; G.~Toker, A.~Gal, J.M.~Eisenberg, Nucl. Phys. A {\bf 362} (1981) 
405.  

\bibitem{CSO70} G. Coremans, et al., Nucl. Phys. B {\bf 16} (1970) 209. 

\bibitem{KSW76} G. Keyes, J. Sacton, J.H. Wickens, M.M. Block, Nuovo Cimento 
A {\bf 31} (1976) 401. 

\bibitem{YAk02} T.~Yamazaki, Y.~Akaishi, Phys. Lett. B {\bf 535} (2002) 70. 

\bibitem{DHW08} A.~Dot\'{e}, T.~Hyodo, W.~Weise, Nucl. Phys. A {\bf 804} 
(2008) 197. 

\bibitem{SGM07} N.V.~Shevchenko, A.~Gal, J.~Mare\v{s}, Phys. Rev. Lett. 
{\bf 98} (2007) 082301; N.V.~Shevchenko, A.~Gal, J.~Mare\v{s}, J.~R\'{e}vai, 
Phys. Rev. C {\bf 76} (2007) 044004. 

\bibitem{ISa07} Y.~Ikeda, T.~Sato, Phys. Rev. C {\bf 76} (2007) 035203. 

\bibitem{GFG08} D.~Gazda, E.~Friedman, A.~Gal, J.~Mare\v{s}, Phys. Rev. C 
{\bf 77} (2008) 045206. 

\bibitem{SNR93} M.M.~Sharma, M.A.~Nagarajan, P.~Ring, Phys. Lett. B {\bf 312} 
(1993) 377. 

\bibitem{CLS06} C.H.~Llewellyn-Smith, remarks made at the Dalitz memorial 
meeting (http://www-thphys.physics.ox.ac.uk/user/SubirSarkar/dalitzmeeting/) 
University of Oxford, 3 June 2006. 

\end{thebibliography}
\end{document}